\documentclass[aps,prl,reprint,groupedaddress]{revtex4-1}
\pdfoutput=1
\usepackage{graphicx}
\usepackage{color}

\newcommand{\beq}{\begin{equation}}
\newcommand{\eeq}{\end{equation}}
\newcommand{\bea}{\begin{eqnarray}}
\newcommand{\eea}{\end{eqnarray}}
\begin{document}
\title{The QCD equation of state at finite density from analytical continuation }


\author{
J.~G\"unther$^{1}$,
R. Bellwied$^{5}$,
S. Borsanyi$^{1}$,
Z. Fodor$^{1,2,3}$,
S. D. Katz$^{2,4}$, 
A. Pasztor$^{1}$, 
C. Ratti$^{5}$
}
\affiliation{$^1$ \small{\it Department of Physics, University of Wuppertal, Gaussstr. 20, D-42119 Wuppertal, Germany}\\
$^2$ \small{\it Inst. for Theoretical Physics, E\"otv\"os University,}\\
\small{\it P\'azm\'any P. s\'et\'any 1/A, H-1117 Budapest, Hungary}\\
$^3$ \small{\it J\"ulich Supercomputing Centre, Forschungszentrum J\"ulich, D-52425
J\"ulich, Germany}\\
$^4$ \small{\it MTA-ELTE "Lend\"ulet" Lattice Gauge Theory Research Group,}\\
\small{\it P\'azm\'any P. s\'et\'any 1/A, H-1117 Budapest, Hungary}\\
$^5$ \small{\it Department of Physics, University of Houston, Houston, TX 77204, USA}\\}

\date{\today}

\begin{abstract}
We determine the equation of state of QCD at finite chemical potential, to
order $(\mu_B/T)^6$, for a system of 2+1 quark flavors. The simulations are
performed at the physical mass for the light and strange quarks on several
lattice spacings; the results are continuum extrapolated using lattices 
of up to $N_t=16$ temporal resolution. The QCD pressure and
interaction measure are calculated along the isentropic trajectories in the
$(T,~\mu_B)$ plane corresponding to the RHIC Beam Energy Scan collision
energies. Their behavior is determined through analytic continuation from
imaginary chemical potentials of the baryonic density. We also determine
the Taylor expansion coefficients around $\mu_B=0$ from the simulations at imaginary chemical potentials.
Strangeness neutrality and charge conservation are imposed,
to match the experimental conditions.
\end{abstract}

\maketitle
The Beam Energy Scan performed at the Relativistic Heavy Ion Collider (RHIC)
constitutes a major effort to explore the phase diagram of QCD at finite
density: by decreasing the collision energy, the net baryonic density of the
system created in the collision can be increased. This allows to scan the phase
diagram of QCD through a change in the initial temperatures and densities.

The experimental program is supplemented by a steady theoretical effort, to 
provide an understanding and a realistic description of the data.
Lattice simulations are one of the main theoretical tools to study the QCD
phase diagram. However, the sign problem limits their range of applicability to
relatively small chemical potentials; the main methods which have been proposed
to circumvent this problem are: multi-parameter reweighting techniques
\cite{Fodor:2001au,Fodor:2001pe,Csikor:2004ik,Fodor:2004nz}, Taylor expansion
of the thermodynamic observables around $\mu_B=0$
\cite{Allton:2002zi,Allton:2005gk,Gavai:2008zr,Basak:2009uv,Kaczmarek:2011zz},
analytical continuation from imaginary chemical potentials
\cite{deForcrand:2002hgr,D'Elia:2002gd,Wu:2006su,D'Elia:2007ke,Conradi:2007be,deForcrand:2008vr,D'Elia:2009tm,Moscicki:2009id}
and the density of state method: \cite{Fodor:2007vv}.
More recent approaches are represented by the use of dual variables
\cite{Gattringer:2014nxa} , and the complex Langevin equation
\cite{Seiler:2012wz,Sexty:2013ica}. However, their application to QCD with
physical  parameters  and  controlled  discretization  has  not  yet  been
achieved.

Among the most sought-after observables, the finite-density equation of state
plays a fundamental role in our understanding of QCD, not only because it
serves as an input for any hydrodynamic approach to the matter created in heavy
ion experiments, but also because it is the main ingredient in the description
of astronomic objects such as dense stars. Even if these stars involve
temperature and density values which are presently not accessible by
first-principle calculations, any progress on the equation of state from
lattice simulations can help to build a bridge between the low-density, high
temperature region of the phase diagram and the high-density, low temperature
one.

The equation of state of QCD at $\mu_B=0$ is known with good accuracy in the
continuum limit
\cite{Borsanyi:2010cj,Borsanyi:2013bia,Bazavov:2014pvz,Borsanyi:2016ksw}, and
is a standard component of state-of-the art hydrodynamic description of heavy
ion collisions. Extensions to finite chemical potential are presently under
control up to order $(\mu_B/T)^2$ \cite{Borsanyi:2012cr}.
Expansion to order $(\mu_B/T)^4$ was only attempted away from the continuum
limit \cite{Hegde:2014sta}. In this letter we calculate the $(\mu_B/T)^4$ and
$(\mu_B/T)^6$ order for the first time with physical quark masses and
in the continuum limit. This allows a reliable determination of the equation
of state up to $\mu_B/T \simeq 2$, compatible with the RHIC energies
down to $\sqrt{s}=14.5$~GeV.
The main
computational challenge so far was to obtain a precise determination of the
fourth and sixth order Taylor expansion coefficient, $c_4$ and $c_6$, which are
notoriously very noisy \cite{Allton:2005gk,Hegde:2014sta}. In fact, the recent
developments in lattice techniques have lead to much smaller discretization
effects in the pion sector. It turned out that in older works
the very same discretization effects have reduced noise
in these coefficients. The traditional technique finds the
$c_4$ and $c_6$ coefficients from the non-Gaussianity of the fluctuations of
conserved charges. In large simulation volumes, however, the central
limit theorem reduces these below the level of detection.
With physical quark masses, fine and large lattices one must seek for an
alternative technique. 

Here we will show that the analytical continuation of the baryonic density from
imaginary chemical potential allows to obtain $c_6$ in the continuum limit.

We also determine the isentropic trajectories in the $(T,~\mu_B)$ plane, which
the system created in a heavy-ion collision follows if dissipations are
negligible. These trajectories are determined by imposing
that the entropy per particle number $S/N_B$ is conserved during the evolution,
and matches the ones determined at the freeze-out for each collision energy.
The calculation of the ratio $S/N_B$ at the freeze-out values for $T$ and
$\mu_B$, and of the isentropic trajectories corresponding to the value at the
freeze-out, are performed for the first time to order $(\mu_B/T)^6$. Selected
thermodynamic quantities (pressure and interaction measure) are calculated
along these trajectories.

The equation of state at finite density can be written as a Taylor expansion around $\mu_B/T=0$:
\bea
\frac{p(\mu_B)}{T^4}=&&c_0(T)+c_2(T)\left(\frac{\mu_B}{T}\right)^2+c_4(T)\left(\frac{\mu_B}{T}\right)^4
\nonumber\\
&&+c_6(T)\left(\frac{\mu_B}{T}\right)^6+\mathcal{O}(\mu_B^8).
\label{eq:coefficients}
\eea
The Taylor coefficients $c_0,~c_2,~c_4,~c_6$ can be calculated on the lattice.
The continuum extrapolated result for $c_2$ was presented for the first time
in \cite{Borsanyi:2012cr}; $c_4$ was calculated at the physical point
\cite{Hegde:2014sta} but only at finite lattice spacing. Here we will present
continuum extrapolated results, at the physical mass, for all of them. Our
results are obtained by imposing the following conditions: 
\bea
\langle n_S\rangle=0~~~\mathrm{and}~~~\langle n_Q\rangle=0.4\langle n_B\rangle\,.
\label{eq:neutrality}
\eea
The method through which we determine $\mu_S(\mu_B)$ and $\mu_Q(\mu_B)$ to
satisfy these requirements was introduced in \cite{Bellwied:2015rza}: the
simulations are performed at matching strange chemical potentials, such that
$n_S=0$ and $n_Q=0.5n_B$. From the simulation we first calculate the
imaginary densities $n_B, n_Q$ and $n_S$ and use higher derivatives in $\mu_S$
and $\mu_Q$ to extrapolate all used observables to the desired condition
(\ref{eq:neutrality}).  The Taylor coefficients in Eq.~(\ref{eq:coefficients})
are the directional derivatives along the line in the ($\mu_B$,$\mu_S$,$\mu_Q$)
space, set by the condition in Eq.~(\ref{eq:neutrality}), calculated at
$\mu_B=0$.

We simulate at six values of imaginary
chemical potentials $\mu_B^{(j)}= i T\pi j/8$, $j=3,4,5,6,6.5$ and $7$.
The principal quantity that we determine in each simulation point at
zero and imaginary $\mu_B$ is:
\bea
\frac{n}{\mu_BT^2}=\frac{T}{\mu_B}\left.\frac{d(p/T^4)}{d(\mu_B/T)}
\right|_{
\langle n_S\rangle=0, ~\langle n_Q\rangle=0.4\langle n_B\rangle, T=\mathrm{const}
}
\label{eq:density}
\eea
Note that $n$ is related to the baryon number density as $\frac{n}{n_B}=1+0.4 \frac{d\mu_Q}{d\mu_B}$.
The first few terms in its Taylor expansion are: $2 c_2 + 4 c_4(\mu_B/T)^2 + 6
c_6(\mu_B/T)^4$: taking derivatives of $n/(\mu_BT^2)$ with respect to
$\mu_B$, we can therefore obtain the desired Taylor coefficients. It turns out
that this method allows a more precise determination of these quantities,
compared to the direct simulation at $\mu_B=0$.

These data are augmented by a $j=0$ data set for all lattices that we used to
calculate the Taylor coefficients using the standard technique.  We used the
$c_2$ data from the $j=0$ runs to calculate the $\mu_B\to 0$ limit of
$n/(\mu_BT^2)$, which we analyzed along with the other simulations with
imaginary $\mu_B$.  These $\mu_B=0$ simulations and the here used methods for
the generalized quark number susceptibilities have already been described in
Ref.~\cite{Bellwied:2015lba}.

Our continuum extrapolation is based on the following lattices: $40^3\times10$,
$48^3\times12$ and $64^3\times16$, in the 4stout staggered discretization.
We refer to the corresponding data sets by the Euclidean temporal resolution
$N_t$=10, 12 and 16, respectively.
(The $j=6.5$ data set consists of $N_t=10$ and $N_t=12$.)
For the details of the lattice action and ensemble parameters see
Ref.~\cite{Bellwied:2015lba}.  
We note that the action in use has a dynamical charm degree of freedom, i.e. we
actually simulate 2+1+1 flavor ensembles for this paper. In
Eq.~(\ref{eq:density}), however, the charm contribution is deliberately not
added (it would be negligible in our temperature range, anyway), and the charm
quark does not couple to the $\mu_B$ chemical potential in our implementation.
We take $c_0$ from the already published $\mu_B=0$ equation of state
with 2+1 dynamical flavors \cite{Borsanyi:2013bia}.
Thus, the results that we present here refer to the 2+1 flavor theory. 
We have shown in our recent work that the effect of the charm quark
is very small for the temperatures of interest in this paper \cite{Borsanyi:2016ksw}.

For each of the above chemical potentials ($\mu_B\ne0$) we run our simulations
at 16 temperature values between 135 and 220 MeV. For higher temperatures,
$T>T_{\rm connect}$, we restrict our study to $c_2$ and $c_4$ based on the data
at $\mu_B=0$. This is justified because, as we will see, $c_6$ is consistent
with zero for $T\geq200$~MeV.  We consider two values for this temperature cut:
$T_{\rm connect} = 200$~MeV and $T_{\rm connect} = 220$~MeV.

The runs on the different lattices do not correspond to the exact same values
of the temperature. Therefore we interpolate each set of $n/(\mu_BT^2)$ data
with fixed imaginary $\mu_B/T$ and lattice resolution separately in
temperature using the following four functions:
\bea
A_1(T)&=&a+b T+c/T+d \arctan(e(T-f))
\nonumber\\
A_2(T)&=&a+b T+c/T+d/(1+e(T-f)^g)^{1/g},
\nonumber\\
A_3(T)&=&a+b T+c T^2+d \arctan(e(T-f))
\nonumber\\
A_4(T)&=&a+b T+c T^2+d/(1+e(T-f)^g)^{1/g}.
\label{eq:Afun}
\eea
The functions $A_{1\dots4}$ are strictly meant as interpolation, the range of
validity obviously cannot extend to very high or very low temperatures. 
For $T>T_{\rm connect}$ we use the $\mu_B=0$ simulations to
extrapolate $n/\mu_BT^2$ to each simulated imaginary $\mu_B/T$ parameter.  This
way we can use the functions (\ref{eq:Afun}) in the full temperature range.

The interpolation is necessary to align the data points to the same temperature
for a given scale setting definition, and to enable us to take the derivative
with respect to the temperature itself, which is needed for the entropy and
energy density. This introduces a correlation between the data at different
temperatures: for this reason, from this point on we analyze the different
temperatures separately. We consider two different scale settings, by fixing
$f_\pi$ and $w_0$ to their physical values. 
The scale setting procedure, and details on the $\mu=0$, $T>0$ ensembles 
are given in Ref.~\cite{Bellwied:2015lba}.

We fit the $\mu_B^2$-dependence of $n(\mu_B,T)/\mu_BT^2$  with three functions:
\bea
B_1(\hat\mu)&=&a+b {\hat\mu}^2+c {\hat\mu}^4\nonumber\\
B_2(\hat\mu)&=&(a+b {\hat\mu}^2)/(1+c {\hat\mu}^2)\nonumber\\
B_3(\hat\mu)&=&a+b {\hat\mu}^2+c \sin({\hat\mu})/{\hat\mu}
\label{fitmu}
\eea
where $\hat\mu=i\mu_B/T$ is a real parameter in our simulations. The first
two functions, polynomial and Pad\`e, are taken as two natural choices when no
prior information on the physics is available.  They are two extremes in the
sense, that $c_8/c_6$ is zero in $B_1$, but large in $B_2$. The third function
reflects our physical expectations at very low and very high temperatures:
below the transition the Hadron Resonance Gas picture predicts that the
$\mu_B$-dependent part of the free energy is proportional to $\cosh(\mu_B/T)$,
which translates to $\sin(\hat\mu)$ for the imaginary density. On the other
hand, at infinite temperature the $a$ and $b$ coefficients exactly describe the
physics.  

\begin{figure}[h]
\includegraphics[width=3in]{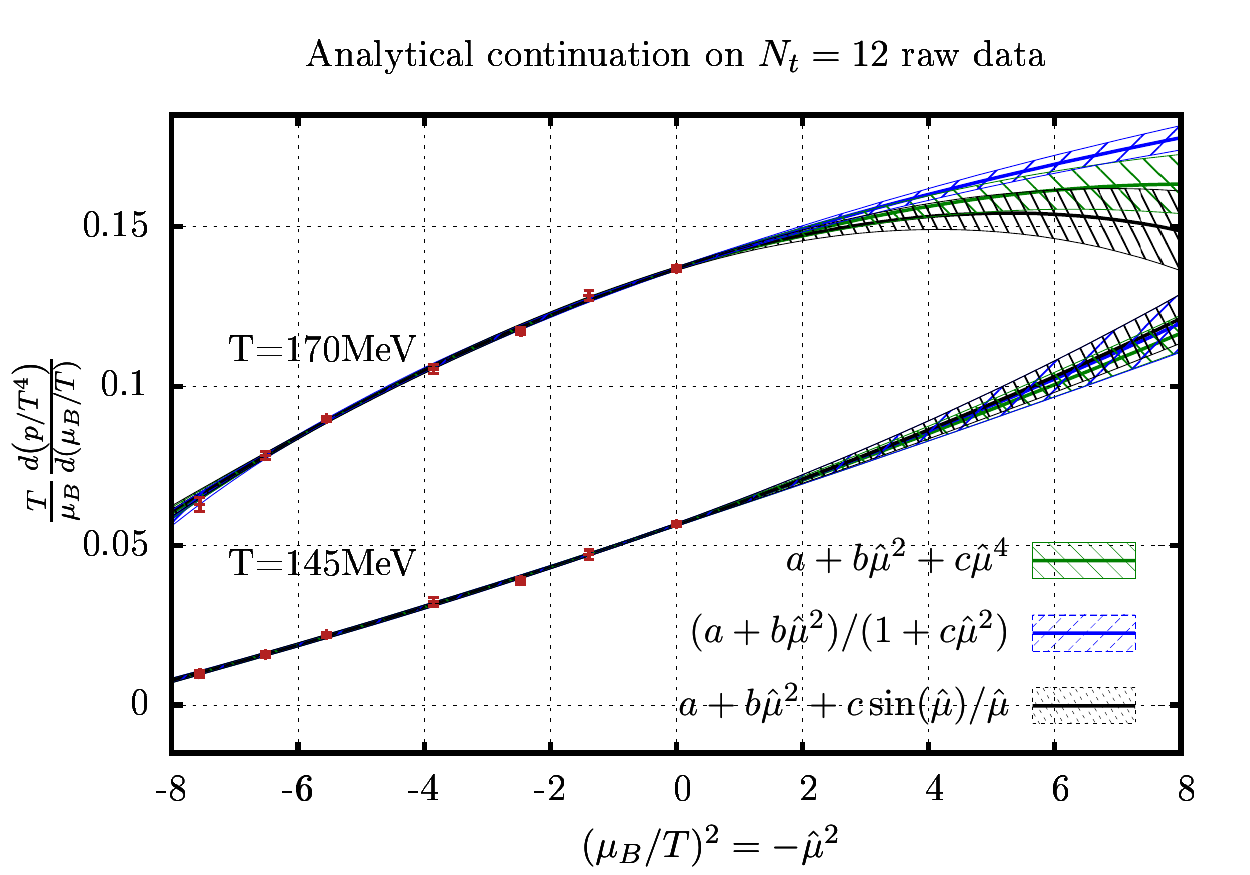}
\caption{
Analytical continuation of $n/(\mu_BT^2)$ from negative to positive
$(\mu_B/T)^2$ for $T=145$ MeV (lower curves) and $T=170$ MeV (upper curves) and
$N_t=12$. The different colors correspond to different fitting functions from
Eqs. (\ref{fitmu}). 
The curves intersect $\mu_B=0$ at $2c_2(T)$, the slope and curvature at
$\mu_B=0$ give $c_4$ and $c_6$, respectively. The plots suggests, that
$c_6$ is negative above $T_c$, and, with marginal statistical significance, 
it is positive below $T_c$. 
\label{fig1}}
\end{figure}

The extrapolation to real $\mu_B$ is performed from the above functional forms:
an example is shown in Fig. \ref{fig1}, where we show the analytical
continuation of $n/(\mu_BT^2)$ from negative to positive $(\mu_B/T)^2$ for
two different temperatures, and $N_t=12$. This plot illustrates the challenge
of analytical continuation: several functions describe the data on the $\mu_B^2<0$
side equally well, but they differ for $\mu_B^2>0$. The systematic error,
that we estimate (among other effects) by varying the fit function, is similar
in size to the statistical uncertainty (calculated using the bootstrap method).

We introduce two options for the continuum extrapolation:
In the first round we fit the results corresponding to the three $N_t$
separately and extract the $a,~b,~c$ coefficients for each one of them. We then
obtain their continuum limit from a linear fit in $1/N_t^2$.
The second option is to combine the fitting of the $\mu_B$-dependence with the
continuum extrapolation.  Consider e.g. the $B_1$ function above. We write it as:
\bea
 B_1(\hat\mu;N_t)&=&a_1+a_2/N_t^2 + b_1 {\hat\mu}^2+b_2 {\hat\mu}^2/N_t^2 
 \nonumber\\
 &&+ c_1 {\hat\mu}^4 + c_2 {\hat\mu}^4/N_t^2.
\eea
In this way we obtain the continuum limit and the $\mu_B$-dependence fit simultaneously.

In total we have 96 different analyses, each one of which produces acceptable
fits; the width of their distribution (with uniform weights) is used as
systematic error.
For each shown quantity we give
the combined error, namely the statistical and systematic errors
added in quadrature.
\begin{figure}[t]
\begin{minipage}{0.48\textwidth}
 \scalebox{.5}{
 \includegraphics{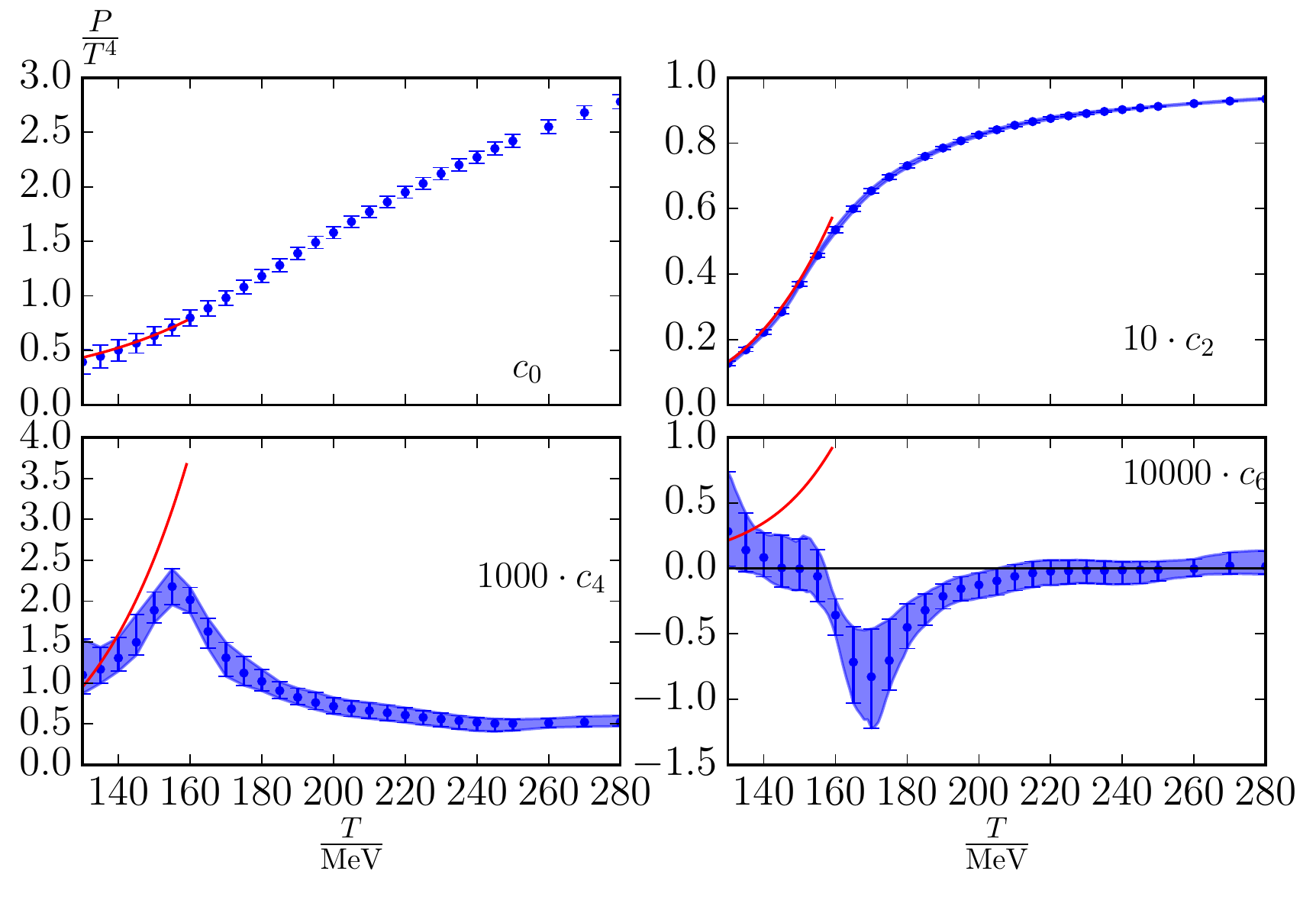}}
 \end{minipage}
\caption{Coefficients $c_0,...c_6$ for the Taylor expansion of the pressure
around $\mu_B=0$. The data are continuum extrapolated and they are presented as
functions of the temperature along with the HRG prediction (red
lines).\label{fig2}} \end{figure}

In Figure \ref{fig2} we show the temperature dependence of the Taylor expansion
coefficients $c_0,...,c_6$ in the continuum limit for the pressure.
Here $c_0$ is taken from our earlier $\mu=0$ work \cite{Borsanyi:2013bia}.
Notice that $c_6$ presents a dip
at a temperature slightly larger than the QCD transition temperature, as
predicted by chiral models \cite{Friman:2011pf}. Besides, $c_6$ is compatible
with zero at $T\geq200$, which makes the use of the Taylor expansion above this
temperature justified. The non-vanishing error on $c_6$ above 220~MeV shows
the intrinsic systematic error of our approach, that includes the use of
the temperature fit Eq.~(\ref{eq:Afun}) and the $\mu_B/T$ functions
$B_{1\dots3}$ in the entire temperature range.

The other quantities are related to the pressure by thermodynamic identities.
The energy density is defined as $\varepsilon= Ts -p +\sum_i\mu_i n_i$, where
$s=[T^4 \partial/\partial T + 4T^3](p/T^4)$ is the entropy density and $i=B,Q$.
The $i=S$ term could be dropped because of the strangeness neutrality
condition.  We used the corresponding $\mu_B,\mu_Q,n_B$ and $n_Q$ values at
each given simulation point. For the $T$-derivative in the entropy density we
used the derivatives of the already fitted functions~(\ref{eq:Afun}). The naive
$T$ derivative of these fit functions is a directional derivative along
constant $\mu_B/T$ and variable $\mu_Q$ and $\mu_S$ defined by
Eq.~(\ref{eq:neutrality}). Using the temperature dependence of $\mu_Q$ one can
calculate the partial $T$-derivative that defines the entropy. The terms in
$\varepsilon$ and $s$ that are related to the variable $\mu_Q/T$ in a
fixed-$\mu_B/T$ dataset are smaller than the overall error. Nevertheless, in
the numerical analysis none of the terms were dropped.

\begin{figure}[h]
 \includegraphics[width=3in]{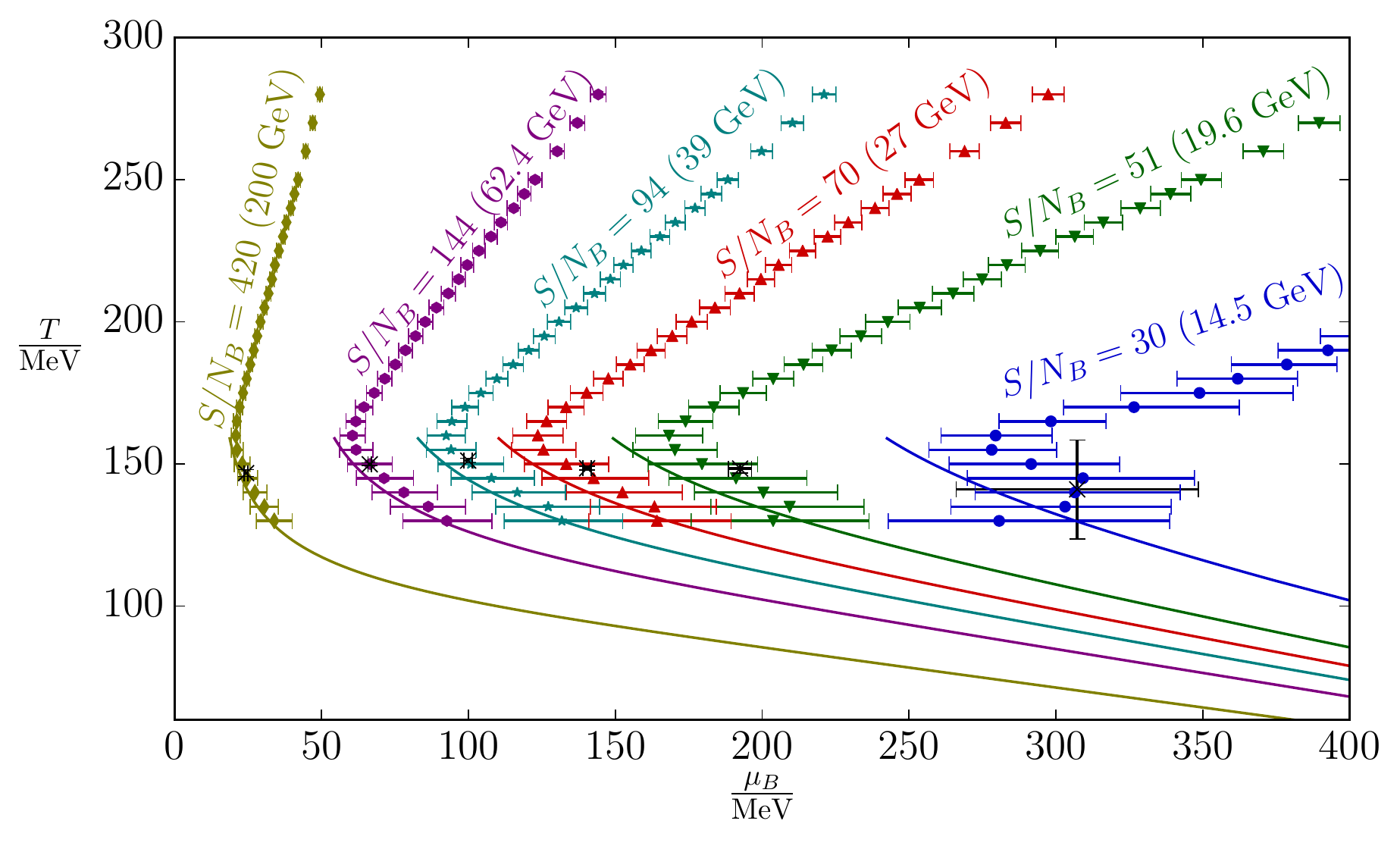}
\caption{The QCD phase diagram in the $(T,~\mu_B)$ plane with the isentropic
trajectories: the contours with fixed $S/N_B$ value. The green points are the
chemical freeze-out parameters extracted in Ref.~\cite{Alba:2014eba}. The
$S/N_B$ ratios correspond to the RHIC energies 200, 62.4, 39, 27, 19.6 and
14.5~GeV. The last point is based on preliminary
STAR data \cite{Luo:2015doi}. The freeze-out parameters are
obtained by a combined fit of net-electric charge and net-proton fluctuations
in the HRG model.\label{fig3}}
\end{figure}

Therefore it is possible to obtain all the thermodynamic quantities
at finite chemical potential. In particular, we start with the entropy density
$s$ and baryonic density $n_B$. These quantities are relevant because, in the
absence of dissipative effects, the medium created in a heavy ion collision
expands without generation of entropy ($S$) and with a fixed baryon number
($N_B$), so that $S/N_B=s/n_B$ is fixed in this case. We calculate the ratio
$s/n_B$ for the values of the freeze-out temperatures and chemical potentials
extracted in Ref. \cite{Alba:2014eba}, which correspond to the various
collision energies of the RHIC beam energy scan. After the initial collision,
the system starts from a point in the $(T,~\mu_B)$ plane and follows a
trajectory which will bring it to one of the freeze-out points. We start from
the freeze-out points and reconstruct the isentropic trajectories backwards in
the $(T,~\mu_B)$ plane. This is done for the first time from lattice QCD
simulations to order $\mu_B^6$. Such isentropic trajectories are shown in Fig.
\ref{fig3}. The black points are the freeze-out parameters from Ref.
\cite{Alba:2014eba}. The last point corresponds to the preliminary analysis of
the new STAR run at 14.5~GeV \cite{Luo:2015doi}. The curves are continued in
the hadronic phase by means of the Hadron Resonance Gas (HRG) model.

\begin{figure}[h]
 \scalebox{.48}{\includegraphics{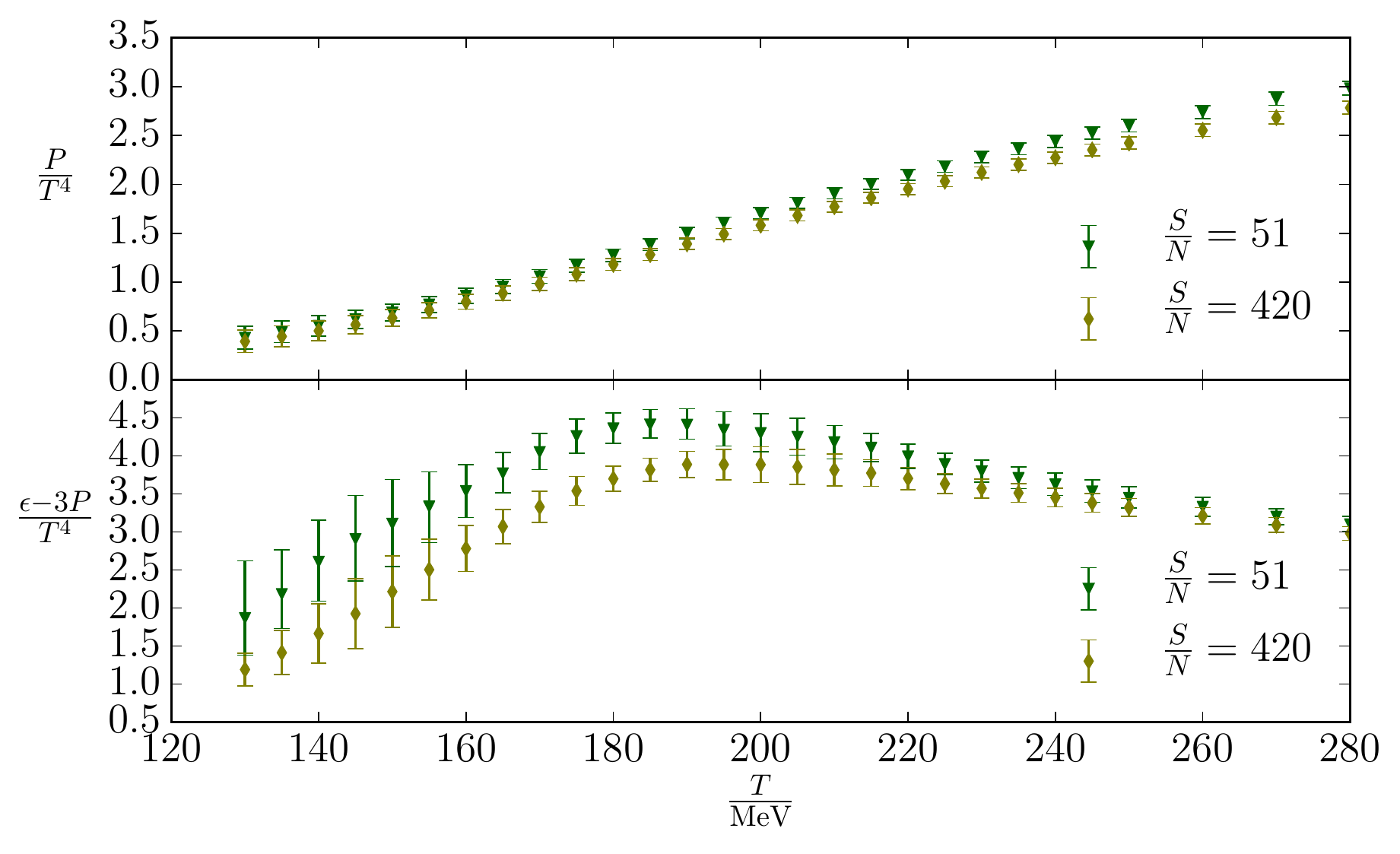}}
\caption{Pressure (upper panel) and interaction measure (lower panel) as functions of temperature, calculated along the highest and lowest isentropic trajectories from Fig. \ref{fig3}.  \label{fig4}}
\end{figure}

We use the continuum extrapolated fit parameters and the formulas in
Eq.~(\ref{fitmu}) to extrapolate the
pressure and the trace anomaly to finite density. In Fig.~\ref{fig4} we plot
these observables for two of the RHIC energies along the isentropic
trajectories of Fig.~\ref{fig3}. 
The effect of the finite chemical potential is more
prominent at high temperature for the pressure, while the interaction measure
is mildly affected by the change in $\mu_B$, and mainly at low temperatures.

In conclusion, we have presented lattice QCD results for the Taylor expansion
coefficients of the pressure up to order $(\mu_B/T)^6$. These
results, simulated at the physical mass and continuum extrapolated, are
achieved for the first time in this paper, using to the method of analytical
continuation of the baryonic density from imaginary chemical potential and
taking its derivatives with respect to $\mu_B$. As our results indicate, this
approach leads to a more precise determination of the coefficients, as compared
to their direct simulation at $\mu_B=0$. Starting from the freeze-out
parameters of Ref.  \cite{Alba:2014eba}, we have then determined the isentropic
trajectories in the $(T,\mu_B)$ plane up to order $(\mu_B/T)^6$, and calculated
the pressure and interaction measure along these trajectories. The results
presented here allow to reliably extend the calculations of the thermodynamic
quantities up to $\mu_B/T\simeq2$, which covers most of the Beam Energy Scan
program at RHIC.

\section*{Acknowledgements}
C.R. would like to thank Volker Koch, Jacquelyn Noronha-Hostler, Jorge Noronha
and Bjorn Schenke for fruitful discussions.  This project was funded by the DFG
grant SFB/TR55. 
This material is based upon work supported by the National Science Foundation
through grant number NSF PHY-1513864 and by the  U.S. Department of Energy,
Office of Science, Office of Nuclear Physics, within the framework
of the Beam Energy Scan Theory (BEST) Topical Collaboration.
An  award  of  computer  time  was  provided by the INCITE program.
This research used resources of the Argonne Leadership Computing Facility,
which is a DOE Office of Science User Facility supported under Contract
DE-AC02-06CH11357.  The authors gratefully acknowledge the Gauss Centre for
Supercomputing (GCS) for providing computing time for a GCS Large-Scale Project
on the GCS share of the supercomputer JUQUEEN \cite{juqueen} at J\"ulich
Supercomputing Centre (JSC).
\bibliography{biblio}

\begin{thebibliography}{33}%
\makeatletter
\providecommand \@ifxundefined [1]{%
 \@ifx{#1\undefined}
}%
\providecommand \@ifnum [1]{%
 \ifnum #1\expandafter \@firstoftwo
 \else \expandafter \@secondoftwo
 \fi
}%
\providecommand \@ifx [1]{%
 \ifx #1\expandafter \@firstoftwo
 \else \expandafter \@secondoftwo
 \fi
}%
\providecommand \natexlab [1]{#1}%
\providecommand \enquote  [1]{``#1''}%
\providecommand \bibnamefont  [1]{#1}%
\providecommand \bibfnamefont [1]{#1}%
\providecommand \citenamefont [1]{#1}%
\providecommand \href@noop [0]{\@secondoftwo}%
\providecommand \href [0]{\begingroup \@sanitize@url \@href}%
\providecommand \@href[1]{\@@startlink{#1}\@@href}%
\providecommand \@@href[1]{\endgroup#1\@@endlink}%
\providecommand \@sanitize@url [0]{\catcode `\\12\catcode `\$12\catcode
  `\&12\catcode `\#12\catcode `\^12\catcode `\_12\catcode `\%12\relax}%
\providecommand \@@startlink[1]{}%
\providecommand \@@endlink[0]{}%
\providecommand \url  [0]{\begingroup\@sanitize@url \@url }%
\providecommand \@url [1]{\endgroup\@href {#1}{\urlprefix }}%
\providecommand \urlprefix  [0]{URL }%
\providecommand \Eprint [0]{\href }%
\providecommand \doibase [0]{http://dx.doi.org/}%
\providecommand \selectlanguage [0]{\@gobble}%
\providecommand \bibinfo  [0]{\@secondoftwo}%
\providecommand \bibfield  [0]{\@secondoftwo}%
\providecommand \translation [1]{[#1]}%
\providecommand \BibitemOpen [0]{}%
\providecommand \bibitemStop [0]{}%
\providecommand \bibitemNoStop [0]{.\EOS\space}%
\providecommand \EOS [0]{\spacefactor3000\relax}%
\providecommand \BibitemShut  [1]{\csname bibitem#1\endcsname}%
\let\auto@bib@innerbib\@empty
\bibitem [{\citenamefont {Fodor}\ and\ \citenamefont
  {Katz}(2002{\natexlab{a}})}]{Fodor:2001au}%
  \BibitemOpen
  \bibfield  {author} {\bibinfo {author} {\bibfnamefont {Z.}~\bibnamefont
  {Fodor}}\ and\ \bibinfo {author} {\bibfnamefont {S.~D.}\ \bibnamefont
  {Katz}},\ }\href {\doibase 10.1016/S0370-2693(02)01583-6} {\bibfield
  {journal} {\bibinfo  {journal} {Phys. Lett.}\ }\textbf {\bibinfo {volume}
  {B534}},\ \bibinfo {pages} {87} (\bibinfo {year} {2002}{\natexlab{a}})},\
  \Eprint {http://arxiv.org/abs/hep-lat/0104001} {arXiv:hep-lat/0104001
  [hep-lat]} \BibitemShut {NoStop}%
\bibitem [{\citenamefont {Fodor}\ and\ \citenamefont
  {Katz}(2002{\natexlab{b}})}]{Fodor:2001pe}%
  \BibitemOpen
  \bibfield  {author} {\bibinfo {author} {\bibfnamefont {Z.}~\bibnamefont
  {Fodor}}\ and\ \bibinfo {author} {\bibfnamefont {S.~D.}\ \bibnamefont
  {Katz}},\ }\href {\doibase 10.1088/1126-6708/2002/03/014} {\bibfield
  {journal} {\bibinfo  {journal} {JHEP}\ }\textbf {\bibinfo {volume} {03}},\
  \bibinfo {pages} {014} (\bibinfo {year} {2002}{\natexlab{b}})},\ \Eprint
  {http://arxiv.org/abs/hep-lat/0106002} {arXiv:hep-lat/0106002 [hep-lat]}
  \BibitemShut {NoStop}%
\bibitem [{\citenamefont {Csikor}\ \emph {et~al.}(2004)\citenamefont {Csikor},
  \citenamefont {Egri}, \citenamefont {Fodor}, \citenamefont {Katz},
  \citenamefont {Szabo},\ and\ \citenamefont {Toth}}]{Csikor:2004ik}%
  \BibitemOpen
  \bibfield  {author} {\bibinfo {author} {\bibfnamefont {F.}~\bibnamefont
  {Csikor}}, \bibinfo {author} {\bibfnamefont {G.~I.}\ \bibnamefont {Egri}},
  \bibinfo {author} {\bibfnamefont {Z.}~\bibnamefont {Fodor}}, \bibinfo
  {author} {\bibfnamefont {S.~D.}\ \bibnamefont {Katz}}, \bibinfo {author}
  {\bibfnamefont {K.~K.}\ \bibnamefont {Szabo}}, \ and\ \bibinfo {author}
  {\bibfnamefont {A.~I.}\ \bibnamefont {Toth}},\ }\href {\doibase
  10.1088/1126-6708/2004/05/046} {\bibfield  {journal} {\bibinfo  {journal}
  {JHEP}\ }\textbf {\bibinfo {volume} {05}},\ \bibinfo {pages} {046} (\bibinfo
  {year} {2004})},\ \Eprint {http://arxiv.org/abs/hep-lat/0401016}
  {arXiv:hep-lat/0401016 [hep-lat]} \BibitemShut {NoStop}%
\bibitem [{\citenamefont {Fodor}\ and\ \citenamefont
  {Katz}(2004)}]{Fodor:2004nz}%
  \BibitemOpen
  \bibfield  {author} {\bibinfo {author} {\bibfnamefont {Z.}~\bibnamefont
  {Fodor}}\ and\ \bibinfo {author} {\bibfnamefont {S.~D.}\ \bibnamefont
  {Katz}},\ }\href {\doibase 10.1088/1126-6708/2004/04/050} {\bibfield
  {journal} {\bibinfo  {journal} {JHEP}\ }\textbf {\bibinfo {volume} {04}},\
  \bibinfo {pages} {050} (\bibinfo {year} {2004})},\ \Eprint
  {http://arxiv.org/abs/hep-lat/0402006} {arXiv:hep-lat/0402006 [hep-lat]}
  \BibitemShut {NoStop}%
\bibitem [{\citenamefont {Allton}\ \emph {et~al.}(2002)\citenamefont {Allton},
  \citenamefont {Ejiri}, \citenamefont {Hands}, \citenamefont {Kaczmarek},
  \citenamefont {Karsch}, \citenamefont {Laermann}, \citenamefont {Schmidt},\
  and\ \citenamefont {Scorzato}}]{Allton:2002zi}%
  \BibitemOpen
  \bibfield  {author} {\bibinfo {author} {\bibfnamefont {C.~R.}\ \bibnamefont
  {Allton}}, \bibinfo {author} {\bibfnamefont {S.}~\bibnamefont {Ejiri}},
  \bibinfo {author} {\bibfnamefont {S.~J.}\ \bibnamefont {Hands}}, \bibinfo
  {author} {\bibfnamefont {O.}~\bibnamefont {Kaczmarek}}, \bibinfo {author}
  {\bibfnamefont {F.}~\bibnamefont {Karsch}}, \bibinfo {author} {\bibfnamefont
  {E.}~\bibnamefont {Laermann}}, \bibinfo {author} {\bibfnamefont
  {C.}~\bibnamefont {Schmidt}}, \ and\ \bibinfo {author} {\bibfnamefont
  {L.}~\bibnamefont {Scorzato}},\ }\href {\doibase 10.1103/PhysRevD.66.074507}
  {\bibfield  {journal} {\bibinfo  {journal} {Phys. Rev.}\ }\textbf {\bibinfo
  {volume} {D66}},\ \bibinfo {pages} {074507} (\bibinfo {year} {2002})},\
  \Eprint {http://arxiv.org/abs/hep-lat/0204010} {arXiv:hep-lat/0204010
  [hep-lat]} \BibitemShut {NoStop}%
\bibitem [{\citenamefont {Allton}\ \emph {et~al.}(2005)\citenamefont {Allton},
  \citenamefont {Doring}, \citenamefont {Ejiri}, \citenamefont {Hands},
  \citenamefont {Kaczmarek}, \citenamefont {Karsch}, \citenamefont {Laermann},\
  and\ \citenamefont {Redlich}}]{Allton:2005gk}%
  \BibitemOpen
  \bibfield  {author} {\bibinfo {author} {\bibfnamefont {C.~R.}\ \bibnamefont
  {Allton}}, \bibinfo {author} {\bibfnamefont {M.}~\bibnamefont {Doring}},
  \bibinfo {author} {\bibfnamefont {S.}~\bibnamefont {Ejiri}}, \bibinfo
  {author} {\bibfnamefont {S.~J.}\ \bibnamefont {Hands}}, \bibinfo {author}
  {\bibfnamefont {O.}~\bibnamefont {Kaczmarek}}, \bibinfo {author}
  {\bibfnamefont {F.}~\bibnamefont {Karsch}}, \bibinfo {author} {\bibfnamefont
  {E.}~\bibnamefont {Laermann}}, \ and\ \bibinfo {author} {\bibfnamefont
  {K.}~\bibnamefont {Redlich}},\ }\href {\doibase 10.1103/PhysRevD.71.054508}
  {\bibfield  {journal} {\bibinfo  {journal} {Phys. Rev.}\ }\textbf {\bibinfo
  {volume} {D71}},\ \bibinfo {pages} {054508} (\bibinfo {year} {2005})},\
  \Eprint {http://arxiv.org/abs/hep-lat/0501030} {arXiv:hep-lat/0501030
  [hep-lat]} \BibitemShut {NoStop}%
\bibitem [{\citenamefont {Gavai}\ and\ \citenamefont
  {Gupta}(2008)}]{Gavai:2008zr}%
  \BibitemOpen
  \bibfield  {author} {\bibinfo {author} {\bibfnamefont {R.~V.}\ \bibnamefont
  {Gavai}}\ and\ \bibinfo {author} {\bibfnamefont {S.}~\bibnamefont {Gupta}},\
  }\href {\doibase 10.1103/PhysRevD.78.114503} {\bibfield  {journal} {\bibinfo
  {journal} {Phys. Rev.}\ }\textbf {\bibinfo {volume} {D78}},\ \bibinfo {pages}
  {114503} (\bibinfo {year} {2008})},\ \Eprint {http://arxiv.org/abs/0806.2233}
  {arXiv:0806.2233 [hep-lat]} \BibitemShut {NoStop}%
\bibitem [{\citenamefont {Basak}\ \emph {et~al.}(2008)\citenamefont {Basak}
  \emph {et~al.}}]{Basak:2009uv}%
  \BibitemOpen
  \bibfield  {author} {\bibinfo {author} {\bibfnamefont {S.}~\bibnamefont
  {Basak}} \emph {et~al.} (\bibinfo {collaboration} {MILC}),\ }\bibfield
  {booktitle} {\emph {\bibinfo {booktitle} {{Proceedings, 26th International
  Symposium on Lattice field theory (Lattice 2008)}}},\ }\href@noop {}
  {\bibfield  {journal} {\bibinfo  {journal} {PoS}\ }\textbf {\bibinfo {volume}
  {LATTICE2008}},\ \bibinfo {pages} {171} (\bibinfo {year} {2008})},\ \Eprint
  {http://arxiv.org/abs/0910.0276} {arXiv:0910.0276 [hep-lat]} \BibitemShut
  {NoStop}%
\bibitem [{\citenamefont {Kaczmarek}\ \emph {et~al.}(2011)\citenamefont
  {Kaczmarek}, \citenamefont {Karsch}, \citenamefont {Laermann}, \citenamefont
  {Miao}, \citenamefont {Mukherjee}, \citenamefont {Petreczky}, \citenamefont
  {Schmidt}, \citenamefont {Soeldner},\ and\ \citenamefont
  {Unger}}]{Kaczmarek:2011zz}%
  \BibitemOpen
  \bibfield  {author} {\bibinfo {author} {\bibfnamefont {O.}~\bibnamefont
  {Kaczmarek}}, \bibinfo {author} {\bibfnamefont {F.}~\bibnamefont {Karsch}},
  \bibinfo {author} {\bibfnamefont {E.}~\bibnamefont {Laermann}}, \bibinfo
  {author} {\bibfnamefont {C.}~\bibnamefont {Miao}}, \bibinfo {author}
  {\bibfnamefont {S.}~\bibnamefont {Mukherjee}}, \bibinfo {author}
  {\bibfnamefont {P.}~\bibnamefont {Petreczky}}, \bibinfo {author}
  {\bibfnamefont {C.}~\bibnamefont {Schmidt}}, \bibinfo {author} {\bibfnamefont
  {W.}~\bibnamefont {Soeldner}}, \ and\ \bibinfo {author} {\bibfnamefont
  {W.}~\bibnamefont {Unger}},\ }\href {\doibase 10.1103/PhysRevD.83.014504}
  {\bibfield  {journal} {\bibinfo  {journal} {Phys. Rev.}\ }\textbf {\bibinfo
  {volume} {D83}},\ \bibinfo {pages} {014504} (\bibinfo {year} {2011})},\
  \Eprint {http://arxiv.org/abs/1011.3130} {arXiv:1011.3130 [hep-lat]}
  \BibitemShut {NoStop}%
\bibitem [{\citenamefont {de~Forcrand}\ and\ \citenamefont
  {Philipsen}(2002)}]{deForcrand:2002hgr}%
  \BibitemOpen
  \bibfield  {author} {\bibinfo {author} {\bibfnamefont {P.}~\bibnamefont
  {de~Forcrand}}\ and\ \bibinfo {author} {\bibfnamefont {O.}~\bibnamefont
  {Philipsen}},\ }\href {\doibase 10.1016/S0550-3213(02)00626-0} {\bibfield
  {journal} {\bibinfo  {journal} {Nucl. Phys.}\ }\textbf {\bibinfo {volume}
  {B642}},\ \bibinfo {pages} {290} (\bibinfo {year} {2002})},\ \Eprint
  {http://arxiv.org/abs/hep-lat/0205016} {arXiv:hep-lat/0205016 [hep-lat]}
  \BibitemShut {NoStop}%
\bibitem [{\citenamefont {D'Elia}\ and\ \citenamefont
  {Lombardo}(2003)}]{D'Elia:2002gd}%
  \BibitemOpen
  \bibfield  {author} {\bibinfo {author} {\bibfnamefont {M.}~\bibnamefont
  {D'Elia}}\ and\ \bibinfo {author} {\bibfnamefont {M.-P.}\ \bibnamefont
  {Lombardo}},\ }\href {\doibase 10.1103/PhysRevD.67.014505} {\bibfield
  {journal} {\bibinfo  {journal} {Phys. Rev.}\ }\textbf {\bibinfo {volume}
  {D67}},\ \bibinfo {pages} {014505} (\bibinfo {year} {2003})},\ \Eprint
  {http://arxiv.org/abs/hep-lat/0209146} {arXiv:hep-lat/0209146 [hep-lat]}
  \BibitemShut {NoStop}%
\bibitem [{\citenamefont {Wu}\ \emph {et~al.}(2007)\citenamefont {Wu},
  \citenamefont {Luo},\ and\ \citenamefont {Chen}}]{Wu:2006su}%
  \BibitemOpen
  \bibfield  {author} {\bibinfo {author} {\bibfnamefont {L.-K.}\ \bibnamefont
  {Wu}}, \bibinfo {author} {\bibfnamefont {X.-Q.}\ \bibnamefont {Luo}}, \ and\
  \bibinfo {author} {\bibfnamefont {H.-S.}\ \bibnamefont {Chen}},\ }\href
  {\doibase 10.1103/PhysRevD.76.034505} {\bibfield  {journal} {\bibinfo
  {journal} {Phys. Rev.}\ }\textbf {\bibinfo {volume} {D76}},\ \bibinfo {pages}
  {034505} (\bibinfo {year} {2007})},\ \Eprint
  {http://arxiv.org/abs/hep-lat/0611035} {arXiv:hep-lat/0611035 [hep-lat]}
  \BibitemShut {NoStop}%
\bibitem [{\citenamefont {D'Elia}\ \emph {et~al.}(2007)\citenamefont {D'Elia},
  \citenamefont {Di~Renzo},\ and\ \citenamefont {Lombardo}}]{D'Elia:2007ke}%
  \BibitemOpen
  \bibfield  {author} {\bibinfo {author} {\bibfnamefont {M.}~\bibnamefont
  {D'Elia}}, \bibinfo {author} {\bibfnamefont {F.}~\bibnamefont {Di~Renzo}}, \
  and\ \bibinfo {author} {\bibfnamefont {M.~P.}\ \bibnamefont {Lombardo}},\
  }\href {\doibase 10.1103/PhysRevD.76.114509} {\bibfield  {journal} {\bibinfo
  {journal} {Phys. Rev.}\ }\textbf {\bibinfo {volume} {D76}},\ \bibinfo {pages}
  {114509} (\bibinfo {year} {2007})},\ \Eprint {http://arxiv.org/abs/0705.3814}
  {arXiv:0705.3814 [hep-lat]} \BibitemShut {NoStop}%
\bibitem [{\citenamefont {Conradi}\ and\ \citenamefont
  {D'Elia}(2007)}]{Conradi:2007be}%
  \BibitemOpen
  \bibfield  {author} {\bibinfo {author} {\bibfnamefont {S.}~\bibnamefont
  {Conradi}}\ and\ \bibinfo {author} {\bibfnamefont {M.}~\bibnamefont
  {D'Elia}},\ }\href {\doibase 10.1103/PhysRevD.76.074501} {\bibfield
  {journal} {\bibinfo  {journal} {Phys. Rev.}\ }\textbf {\bibinfo {volume}
  {D76}},\ \bibinfo {pages} {074501} (\bibinfo {year} {2007})},\ \Eprint
  {http://arxiv.org/abs/0707.1987} {arXiv:0707.1987 [hep-lat]} \BibitemShut
  {NoStop}%
\bibitem [{\citenamefont {de~Forcrand}\ and\ \citenamefont
  {Philipsen}(2008)}]{deForcrand:2008vr}%
  \BibitemOpen
  \bibfield  {author} {\bibinfo {author} {\bibfnamefont {P.}~\bibnamefont
  {de~Forcrand}}\ and\ \bibinfo {author} {\bibfnamefont {O.}~\bibnamefont
  {Philipsen}},\ }\href {\doibase 10.1088/1126-6708/2008/11/012} {\bibfield
  {journal} {\bibinfo  {journal} {JHEP}\ }\textbf {\bibinfo {volume} {11}},\
  \bibinfo {pages} {012} (\bibinfo {year} {2008})},\ \Eprint
  {http://arxiv.org/abs/0808.1096} {arXiv:0808.1096 [hep-lat]} \BibitemShut
  {NoStop}%
\bibitem [{\citenamefont {D'Elia}\ and\ \citenamefont
  {Sanfilippo}(2009)}]{D'Elia:2009tm}%
  \BibitemOpen
  \bibfield  {author} {\bibinfo {author} {\bibfnamefont {M.}~\bibnamefont
  {D'Elia}}\ and\ \bibinfo {author} {\bibfnamefont {F.}~\bibnamefont
  {Sanfilippo}},\ }\href {\doibase 10.1103/PhysRevD.80.014502} {\bibfield
  {journal} {\bibinfo  {journal} {Phys. Rev.}\ }\textbf {\bibinfo {volume}
  {D80}},\ \bibinfo {pages} {014502} (\bibinfo {year} {2009})},\ \Eprint
  {http://arxiv.org/abs/0904.1400} {arXiv:0904.1400 [hep-lat]} \BibitemShut
  {NoStop}%
\bibitem [{\citenamefont {Moscicki}\ \emph {et~al.}(2010)\citenamefont
  {Moscicki}, \citenamefont {Wos}, \citenamefont {Lamanna}, \citenamefont
  {de~Forcrand},\ and\ \citenamefont {Philipsen}}]{Moscicki:2009id}%
  \BibitemOpen
  \bibfield  {author} {\bibinfo {author} {\bibfnamefont {J.~T.}\ \bibnamefont
  {Moscicki}}, \bibinfo {author} {\bibfnamefont {M.}~\bibnamefont {Wos}},
  \bibinfo {author} {\bibfnamefont {M.}~\bibnamefont {Lamanna}}, \bibinfo
  {author} {\bibfnamefont {P.}~\bibnamefont {de~Forcrand}}, \ and\ \bibinfo
  {author} {\bibfnamefont {O.}~\bibnamefont {Philipsen}},\ }\href {\doibase
  10.1016/j.cpc.2010.06.027} {\bibfield  {journal} {\bibinfo  {journal}
  {Comput. Phys. Commun.}\ }\textbf {\bibinfo {volume} {181}},\ \bibinfo
  {pages} {1715} (\bibinfo {year} {2010})},\ \Eprint
  {http://arxiv.org/abs/0911.5682} {arXiv:0911.5682 [cs.DC]} \BibitemShut
  {NoStop}%
\bibitem [{\citenamefont {Fodor}\ \emph {et~al.}(2007)\citenamefont {Fodor},
  \citenamefont {Katz},\ and\ \citenamefont {Schmidt}}]{Fodor:2007vv}%
  \BibitemOpen
  \bibfield  {author} {\bibinfo {author} {\bibfnamefont {Z.}~\bibnamefont
  {Fodor}}, \bibinfo {author} {\bibfnamefont {S.~D.}\ \bibnamefont {Katz}}, \
  and\ \bibinfo {author} {\bibfnamefont {C.}~\bibnamefont {Schmidt}},\ }\href
  {\doibase 10.1088/1126-6708/2007/03/121} {\bibfield  {journal} {\bibinfo
  {journal} {JHEP}\ }\textbf {\bibinfo {volume} {0703}},\ \bibinfo {pages}
  {121} (\bibinfo {year} {2007})},\ \Eprint
  {http://arxiv.org/abs/hep-lat/0701022} {arXiv:hep-lat/0701022 [hep-lat]}
  \BibitemShut {NoStop}%
\bibitem [{\citenamefont {Gattringer}(2014)}]{Gattringer:2014nxa}%
  \BibitemOpen
  \bibfield  {author} {\bibinfo {author} {\bibfnamefont {C.}~\bibnamefont
  {Gattringer}},\ }\bibfield  {booktitle} {\emph {\bibinfo {booktitle}
  {{Proceedings, 31st International Symposium on Lattice Field Theory (Lattice
  2013)}}},\ }\href@noop {} {\bibfield  {journal} {\bibinfo  {journal} {PoS}\
  }\textbf {\bibinfo {volume} {LATTICE2013}},\ \bibinfo {pages} {002} (\bibinfo
  {year} {2014})},\ \Eprint {http://arxiv.org/abs/1401.7788} {arXiv:1401.7788
  [hep-lat]} \BibitemShut {NoStop}%
\bibitem [{\citenamefont {Seiler}\ \emph {et~al.}(2013)\citenamefont {Seiler},
  \citenamefont {Sexty},\ and\ \citenamefont {Stamatescu}}]{Seiler:2012wz}%
  \BibitemOpen
  \bibfield  {author} {\bibinfo {author} {\bibfnamefont {E.}~\bibnamefont
  {Seiler}}, \bibinfo {author} {\bibfnamefont {D.}~\bibnamefont {Sexty}}, \
  and\ \bibinfo {author} {\bibfnamefont {I.-O.}\ \bibnamefont {Stamatescu}},\
  }\href {\doibase 10.1016/j.physletb.2013.04.062} {\bibfield  {journal}
  {\bibinfo  {journal} {Phys. Lett.}\ }\textbf {\bibinfo {volume} {B723}},\
  \bibinfo {pages} {213} (\bibinfo {year} {2013})},\ \Eprint
  {http://arxiv.org/abs/1211.3709} {arXiv:1211.3709 [hep-lat]} \BibitemShut
  {NoStop}%
\bibitem [{\citenamefont {Sexty}(2014)}]{Sexty:2013ica}%
  \BibitemOpen
  \bibfield  {author} {\bibinfo {author} {\bibfnamefont {D.}~\bibnamefont
  {Sexty}},\ }\href {\doibase 10.1016/j.physletb.2014.01.019} {\bibfield
  {journal} {\bibinfo  {journal} {Phys. Lett.}\ }\textbf {\bibinfo {volume}
  {B729}},\ \bibinfo {pages} {108} (\bibinfo {year} {2014})},\ \Eprint
  {http://arxiv.org/abs/1307.7748} {arXiv:1307.7748 [hep-lat]} \BibitemShut
  {NoStop}%
\bibitem [{\citenamefont {Borsanyi}\ \emph {et~al.}(2010)\citenamefont
  {Borsanyi}, \citenamefont {Endrodi}, \citenamefont {Fodor}, \citenamefont
  {Jakovac}, \citenamefont {Katz}, \citenamefont {Krieg}, \citenamefont
  {Ratti},\ and\ \citenamefont {Szabo}}]{Borsanyi:2010cj}%
  \BibitemOpen
  \bibfield  {author} {\bibinfo {author} {\bibfnamefont {S.}~\bibnamefont
  {Borsanyi}}, \bibinfo {author} {\bibfnamefont {G.}~\bibnamefont {Endrodi}},
  \bibinfo {author} {\bibfnamefont {Z.}~\bibnamefont {Fodor}}, \bibinfo
  {author} {\bibfnamefont {A.}~\bibnamefont {Jakovac}}, \bibinfo {author}
  {\bibfnamefont {S.~D.}\ \bibnamefont {Katz}}, \bibinfo {author}
  {\bibfnamefont {S.}~\bibnamefont {Krieg}}, \bibinfo {author} {\bibfnamefont
  {C.}~\bibnamefont {Ratti}}, \ and\ \bibinfo {author} {\bibfnamefont {K.~K.}\
  \bibnamefont {Szabo}},\ }\href {\doibase 10.1007/JHEP11(2010)077} {\bibfield
  {journal} {\bibinfo  {journal} {JHEP}\ }\textbf {\bibinfo {volume} {11}},\
  \bibinfo {pages} {077} (\bibinfo {year} {2010})},\ \Eprint
  {http://arxiv.org/abs/1007.2580} {arXiv:1007.2580 [hep-lat]} \BibitemShut
  {NoStop}%
\bibitem [{\citenamefont {Borsanyi}\ \emph {et~al.}(2014)\citenamefont
  {Borsanyi}, \citenamefont {Fodor}, \citenamefont {Hoelbling}, \citenamefont
  {Katz}, \citenamefont {Krieg},\ and\ \citenamefont
  {Szabo}}]{Borsanyi:2013bia}%
  \BibitemOpen
  \bibfield  {author} {\bibinfo {author} {\bibfnamefont {S.}~\bibnamefont
  {Borsanyi}}, \bibinfo {author} {\bibfnamefont {Z.}~\bibnamefont {Fodor}},
  \bibinfo {author} {\bibfnamefont {C.}~\bibnamefont {Hoelbling}}, \bibinfo
  {author} {\bibfnamefont {S.~D.}\ \bibnamefont {Katz}}, \bibinfo {author}
  {\bibfnamefont {S.}~\bibnamefont {Krieg}}, \ and\ \bibinfo {author}
  {\bibfnamefont {K.~K.}\ \bibnamefont {Szabo}},\ }\href {\doibase
  10.1016/j.physletb.2014.01.007} {\bibfield  {journal} {\bibinfo  {journal}
  {Phys. Lett.}\ }\textbf {\bibinfo {volume} {B730}},\ \bibinfo {pages} {99}
  (\bibinfo {year} {2014})},\ \Eprint {http://arxiv.org/abs/1309.5258}
  {arXiv:1309.5258 [hep-lat]} \BibitemShut {NoStop}%
\bibitem [{\citenamefont {Bazavov}\ \emph {et~al.}(2014)\citenamefont {Bazavov}
  \emph {et~al.}}]{Bazavov:2014pvz}%
  \BibitemOpen
  \bibfield  {author} {\bibinfo {author} {\bibfnamefont {A.}~\bibnamefont
  {Bazavov}} \emph {et~al.} (\bibinfo {collaboration} {HotQCD}),\ }\href
  {\doibase 10.1103/PhysRevD.90.094503} {\bibfield  {journal} {\bibinfo
  {journal} {Phys. Rev.}\ }\textbf {\bibinfo {volume} {D90}},\ \bibinfo {pages}
  {094503} (\bibinfo {year} {2014})},\ \Eprint {http://arxiv.org/abs/1407.6387}
  {arXiv:1407.6387 [hep-lat]} \BibitemShut {NoStop}%
\bibitem [{\citenamefont {Borsanyi}\ \emph {et~al.}(2016)\citenamefont
  {Borsanyi} \emph {et~al.}}]{Borsanyi:2016ksw}%
  \BibitemOpen
  \bibfield  {author} {\bibinfo {author} {\bibfnamefont {S.}~\bibnamefont
  {Borsanyi}} \emph {et~al.},\ }\href@noop {} {\  (\bibinfo {year} {2016})},\
  \Eprint {http://arxiv.org/abs/1606.07494} {arXiv:1606.07494 [hep-lat]}
  \BibitemShut {NoStop}%
\bibitem [{\citenamefont {Borsanyi}\ \emph {et~al.}(2012)\citenamefont
  {Borsanyi}, \citenamefont {Endrodi}, \citenamefont {Fodor}, \citenamefont
  {Katz}, \citenamefont {Krieg}, \citenamefont {Ratti},\ and\ \citenamefont
  {Szabo}}]{Borsanyi:2012cr}%
  \BibitemOpen
  \bibfield  {author} {\bibinfo {author} {\bibfnamefont {S.}~\bibnamefont
  {Borsanyi}}, \bibinfo {author} {\bibfnamefont {G.}~\bibnamefont {Endrodi}},
  \bibinfo {author} {\bibfnamefont {Z.}~\bibnamefont {Fodor}}, \bibinfo
  {author} {\bibfnamefont {S.~D.}\ \bibnamefont {Katz}}, \bibinfo {author}
  {\bibfnamefont {S.}~\bibnamefont {Krieg}}, \bibinfo {author} {\bibfnamefont
  {C.}~\bibnamefont {Ratti}}, \ and\ \bibinfo {author} {\bibfnamefont {K.~K.}\
  \bibnamefont {Szabo}},\ }\href {\doibase 10.1007/JHEP08(2012)053} {\bibfield
  {journal} {\bibinfo  {journal} {JHEP}\ }\textbf {\bibinfo {volume} {08}},\
  \bibinfo {pages} {053} (\bibinfo {year} {2012})},\ \Eprint
  {http://arxiv.org/abs/1204.6710} {arXiv:1204.6710 [hep-lat]} \BibitemShut
  {NoStop}%
\bibitem [{\citenamefont {Hegde}(2014)}]{Hegde:2014sta}%
  \BibitemOpen
  \bibfield  {author} {\bibinfo {author} {\bibfnamefont {P.}~\bibnamefont
  {Hegde}} (\bibinfo {collaboration} {BNL--Bielefeld--CCNU}),\ }\href {\doibase
  10.1016/j.nuclphysa.2014.08.089} {\bibfield  {journal} {\bibinfo  {journal}
  {Nucl.Phys.}\ }\textbf {\bibinfo {volume} {A931}},\ \bibinfo {pages} {851}
  (\bibinfo {year} {2014})},\ \Eprint {http://arxiv.org/abs/1408.6305}
  {arXiv:1408.6305 [hep-lat]} \BibitemShut {NoStop}%
\bibitem [{\citenamefont {Bellwied}\ \emph
  {et~al.}(2015{\natexlab{a}})\citenamefont {Bellwied}, \citenamefont
  {Borsanyi}, \citenamefont {Fodor}, \citenamefont {G{\"u}nther}, \citenamefont
  {Katz}, \citenamefont {Ratti},\ and\ \citenamefont
  {Szabo}}]{Bellwied:2015rza}%
  \BibitemOpen
  \bibfield  {author} {\bibinfo {author} {\bibfnamefont {R.}~\bibnamefont
  {Bellwied}}, \bibinfo {author} {\bibfnamefont {S.}~\bibnamefont {Borsanyi}},
  \bibinfo {author} {\bibfnamefont {Z.}~\bibnamefont {Fodor}}, \bibinfo
  {author} {\bibfnamefont {J.}~\bibnamefont {G{\"u}nther}}, \bibinfo {author}
  {\bibfnamefont {S.~D.}\ \bibnamefont {Katz}}, \bibinfo {author}
  {\bibfnamefont {C.}~\bibnamefont {Ratti}}, \ and\ \bibinfo {author}
  {\bibfnamefont {K.~K.}\ \bibnamefont {Szabo}},\ }\href {\doibase
  10.1016/j.physletb.2015.11.011} {\bibfield  {journal} {\bibinfo  {journal}
  {Phys. Lett.}\ }\textbf {\bibinfo {volume} {B751}},\ \bibinfo {pages} {559}
  (\bibinfo {year} {2015}{\natexlab{a}})},\ \Eprint
  {http://arxiv.org/abs/1507.07510} {arXiv:1507.07510 [hep-lat]} \BibitemShut
  {NoStop}%
\bibitem [{\citenamefont {Bellwied}\ \emph
  {et~al.}(2015{\natexlab{b}})\citenamefont {Bellwied}, \citenamefont
  {Borsanyi}, \citenamefont {Fodor}, \citenamefont {Katz}, \citenamefont
  {Pasztor}, \citenamefont {Ratti},\ and\ \citenamefont
  {Szabo}}]{Bellwied:2015lba}%
  \BibitemOpen
  \bibfield  {author} {\bibinfo {author} {\bibfnamefont {R.}~\bibnamefont
  {Bellwied}}, \bibinfo {author} {\bibfnamefont {S.}~\bibnamefont {Borsanyi}},
  \bibinfo {author} {\bibfnamefont {Z.}~\bibnamefont {Fodor}}, \bibinfo
  {author} {\bibfnamefont {S.~D.}\ \bibnamefont {Katz}}, \bibinfo {author}
  {\bibfnamefont {A.}~\bibnamefont {Pasztor}}, \bibinfo {author} {\bibfnamefont
  {C.}~\bibnamefont {Ratti}}, \ and\ \bibinfo {author} {\bibfnamefont {K.~K.}\
  \bibnamefont {Szabo}},\ }\href {\doibase 10.1103/PhysRevD.92.114505}
  {\bibfield  {journal} {\bibinfo  {journal} {Phys. Rev.}\ }\textbf {\bibinfo
  {volume} {D92}},\ \bibinfo {pages} {114505} (\bibinfo {year}
  {2015}{\natexlab{b}})},\ \Eprint {http://arxiv.org/abs/1507.04627}
  {arXiv:1507.04627 [hep-lat]} \BibitemShut {NoStop}%
\bibitem [{\citenamefont {Friman}\ \emph {et~al.}(2011)\citenamefont {Friman},
  \citenamefont {Karsch}, \citenamefont {Redlich},\ and\ \citenamefont
  {Skokov}}]{Friman:2011pf}%
  \BibitemOpen
  \bibfield  {author} {\bibinfo {author} {\bibfnamefont {B.}~\bibnamefont
  {Friman}}, \bibinfo {author} {\bibfnamefont {F.}~\bibnamefont {Karsch}},
  \bibinfo {author} {\bibfnamefont {K.}~\bibnamefont {Redlich}}, \ and\
  \bibinfo {author} {\bibfnamefont {V.}~\bibnamefont {Skokov}},\ }\href
  {\doibase 10.1140/epjc/s10052-011-1694-2} {\bibfield  {journal} {\bibinfo
  {journal} {Eur. Phys. J.}\ }\textbf {\bibinfo {volume} {C71}},\ \bibinfo
  {pages} {1694} (\bibinfo {year} {2011})},\ \Eprint
  {http://arxiv.org/abs/1103.3511} {arXiv:1103.3511 [hep-ph]} \BibitemShut
  {NoStop}%
\bibitem [{\citenamefont {Alba}\ \emph {et~al.}(2014)\citenamefont {Alba},
  \citenamefont {Alberico}, \citenamefont {Bellwied}, \citenamefont {Bluhm},
  \citenamefont {Mantovani~Sarti}, \citenamefont {Nahrgang},\ and\
  \citenamefont {Ratti}}]{Alba:2014eba}%
  \BibitemOpen
  \bibfield  {author} {\bibinfo {author} {\bibfnamefont {P.}~\bibnamefont
  {Alba}}, \bibinfo {author} {\bibfnamefont {W.}~\bibnamefont {Alberico}},
  \bibinfo {author} {\bibfnamefont {R.}~\bibnamefont {Bellwied}}, \bibinfo
  {author} {\bibfnamefont {M.}~\bibnamefont {Bluhm}}, \bibinfo {author}
  {\bibfnamefont {V.}~\bibnamefont {Mantovani~Sarti}}, \bibinfo {author}
  {\bibfnamefont {M.}~\bibnamefont {Nahrgang}}, \ and\ \bibinfo {author}
  {\bibfnamefont {C.}~\bibnamefont {Ratti}},\ }\href {\doibase
  10.1016/j.physletb.2014.09.052} {\bibfield  {journal} {\bibinfo  {journal}
  {Phys. Lett.}\ }\textbf {\bibinfo {volume} {B738}},\ \bibinfo {pages} {305}
  (\bibinfo {year} {2014})},\ \Eprint {http://arxiv.org/abs/1403.4903}
  {arXiv:1403.4903 [hep-ph]} \BibitemShut {NoStop}%
\bibitem [{\citenamefont {Luo}(2015)}]{Luo:2015doi}%
  \BibitemOpen
  \bibfield  {author} {\bibinfo {author} {\bibfnamefont {X.}~\bibnamefont
  {Luo}}\ }(\bibinfo {year} {2015})\ \Eprint {http://arxiv.org/abs/1512.09215}
  {arXiv:1512.09215 [nucl-ex]} \BibitemShut {NoStop}%
\bibitem [{juq(2015)}]{juqueen}%
  \BibitemOpen
  \href@noop {} {\emph {\bibinfo {title} {JUQUEEN: IBM Blue Gene/Q
  Supercomputer System at the J{\"u}lich Supercomputing Centre}}},\ \bibinfo
  {type} {Tech. Rep.}\ \bibinfo {number} {1 A1}\ (\bibinfo  {institution}
  {J{\"u}lich Supercomputing Centre},\ \bibinfo {address}
  {http://dx.doi.org/10.17815/jlsrf-1-18},\ \bibinfo {year} {2015})\BibitemShut
  {NoStop}%
\end{thebibliography}%


%

\end{document}